\documentclass[onecolumn,aps,preprint,amsmath,amssymb]{revtex4}
\usepackage{xcolor}
\usepackage{empheq}	
\usepackage{bm}
\usepackage{dcolumn} 
\usepackage{stmaryrd} 
\begin{document}
\newcommand{\eq}{\begin{equation}}                                                                         
\newcommand{\eqe}{\end{equation}}             

\title{Time-dependent analytic solutions for water waves above sea of varying depths} 
\author{Imre Ferenc Barna$^{1}$   Mih\'aly Andr\'as Pocsai$^{1}$ and L\'aszl\'o M\'aty\'as$^{2}$}
\address{ $^1$ Wigner Research Center for Physics, 
\\ Konkoly-Thege Mikl\'os \'ut 29 - 33, 1121 Budapest, Hungary \\
$^2$  Department of Bioengineering, Faculty of Economics, Socio-Human Sciences
and Engineering, Sapientia Hungarian University of Transylvania,
Libert\u{a}tii sq. 1, 530104 Miercurea Ciuc, Romania} 
 
\begin{abstract} 
We investigate a hydrodynamic equation system 
which - with some approximation - is capable to 
describe the tsunami propagation in the open ocean 
with the time-dependent self-similar Ansatz.  
We found analytic solutions how the wave height and velocity behave 
in time and space for constant and linear seabed functions.  
First we study waves on open water, where the seabed can be considered relatively constant, sufficiently far from 
the shore. In the second part of the study we also consider a seabed which is oblique. Finally, we apply the most common traveling wave Ansatz and present almost trivial solutions as well. 
   
\end{abstract}
\maketitle
\section{Introduction}
Wave propagation in non-linear media is a fascinating field in physics with enormous literature, (without completeness we just mention some 
relevant monographs) \cite{genwave00,genwave0, genwave1, genwave2, genwave3, genwave4, genwave5}.  
Narrowing the scientific question to the dynamics of various waves in sea or fresh water is still an immerse problem with considerable  literature \cite{wave1,wave2,wave3,wave4,wave5,wave6,wave7,wave8,wave9,wave10,wave11}.  
The very first pioneering work was written by Airy in 1841 with the title of "Tides and Waves" \cite{airy}. 
Interaction of water waves with ships \cite{ship1} is also a crucial question both from theoretical and engineering sides as well. 
 
Regarding water waves one may find important numerical and analytical studies of Bussinesq approximation 
 \cite{MaFu2006,Wa2007,Wa2008,RoCh2012, ShKi2012,Wa2012,HeSeZe2014,KaDe2018,KoDi2012,YaMaBa2017}.
 There is also a Boussinesq approximation with dissipative dynamics and possible density variations 
\cite{DaPa2009,GaWiAu2015,An2016,WeHeAh2018, XiZh2006, LaGr2018}. 
Experiments for certain parameter values are also realized \cite{AhHeFuBo2012,AhBoHe2014}.  
Connections related to radiation and environment can be found in \cite{PaEmPr2003}.  

The tragedy of the 2004 December Indian Ocean Tsunami highlighted that investigation of such physical 
and mathematical problems are indispensable and the obtained results can save human lives. Studies on destructive weather phenomena one may find in \cite{Li2021}. 

To study such effects diverse non-linear partial differential equations(PDEs) have to be investigated with various methods. 
Tsunamis are long life non-dispersive waves which can be well described with solitons and with the   
corresponding mathematics. One may also find wave equations with fractional derivatives in \cite{YaBa2014,AbuBa2020}.
 
In the following we choose a completely different path, we investigate the long-time dispersion and decay 
of such kind of fluid equations with the self-similar Ansatz \cite{sedov,zeldovich}. This Ansatz inherently 
contains two exponents - for each dynamical variable - which describes the asymptotic decay and dispersion of 
the solutions. This Ansatz is the natural trial function of the regular diffusion (or heat conduction) equation and gives 
the Gaussian (or fundamental) solution after some easy mathematical steps. 
If the parameter dependences of these solutions are systematically studied and analyzed then a well-established 
physical image emerges in front of us. Of course, additional mathematical 
methods (like using generalized symmetries) also exist to obtain other solutions like those in \cite{cole}.

This study is organically linked to our personal long-term strategy in which we systematically investigate the fundamental hydrodynamic 
systems one after another and analyze with the physically relevant self-similar Ansatz.   
Till now we published more than ten papers - (some of them are  \cite{imre1,imre2,imre3,imre4}) - and two book chapters 
\cite{imre_book, imre_book2} in this field.  
In our last two publications we investigates the question of finding analytic solutions for the rotating and stratified Euler equations \cite{rot}, 
and the analysis of a two-fluid model where the Euler and the Navier-Stokes equations were coupled \cite{two}. 
Analytic solutions of Navier-Stokes equation with varying viscosity with density have been also found \cite{DoZh2021}. 

The new feature of the present study is that we apply different analytic sea-bed functions and present analytic solutions for each of them. 
We successfully applied this investigation method for the KPZ surface growth model \cite{kpz1,kpz2} where half 
dozen different kind of noise terms were considered and analyzed with the self-similar and traveling wave Ans\"atze.     
To the best of our knowledge, there are no such time-dependent self-similar 
solutions known, presented and analyzed in the scientific literature for this water-wave equation system.

\section{Theory and Results}
\begin{figure} 
\scalebox{1.2}{
\rotatebox{0}{\includegraphics{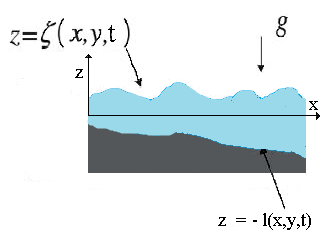}}}
\vspace*{-0.5cm}
\caption{The geometry of the flow where $\zeta(x,y,t)$ is the wave height and $l(x,y,t)$ is the seabed function, respectively.} 
\label{egyes}        
\end{figure}

 In the following we will investigate the next PDE system for water  
waves \cite{wave2} 
 \begin{subequations}\label{eq0:system1}
\begin{empheq} {align}
 \frac{\partial \zeta}{\partial t} +   \frac{\partial }{\partial x} (u\cdot l) +  \frac{\partial }{\partial y} (v\cdot l) & = 0,  \label{eq0:aa1} \\ 
 \frac{\partial}{\partial t} (u \cdot l) +   c^2 \frac{\partial }{\partial x} \zeta & = 0,  \label{eq0:aa2}  \\ 
  \frac{\partial}{\partial t} (v \cdot l) +   c^2 \frac{\partial }{\partial y} \zeta  & = 0,   \label{eq0:aa3}
\end{empheq}
\end{subequations}
where the dynamical variables are the wave height $\zeta(x,y,t)$ and the two orthogonal horizontal fluid velocity components $u(x,y,t), v(x,y,t)$. 
The function $l(x,y)$ or $l(x,y,t) $is the depth of the sea, or the function of the seabed.  
We will investigate both cases, time-independent and time-dependent seabed functions and the differences will be highlighted. 
The mathematical form of the self-similar Ansatz inherently makes this 
comparison possible. 
We consider that water is inviscid, incompressible and irrotational. 
The variable local wave propagation speed is $c = \sqrt{G\cdot l(x,y,t)}$ 
with the $G \approx 10 \>\>  \frac{m}{s^2}  $ gravity acceleration. 
Figure (\ref{egyes}) shows the geometry of the investigated flow problem. 
The wave height of a surface wave $\zeta(x,y,t)$ is the difference between the elevations of a crest and a neighboring trough. 
 The original depth of the water is the distance between the minima of the wave height function $\zeta(x,y,t)$
and the maxima of the negative seabed function $l(x,y)$ (or $l(x,y,t)$) and can be shifted with any arbitrary constant.   
At this point we have to mention that a large number of such equations are derived and 
part of them solved in the work of \cite{wave9}. 
Unfortunately, the solution functions are not analyzed, not visualized on figures and no 
detailed parameter studies were presented which would be desirable for physicist or engineers.  Water waves 
on variable depth is also an exhaustively investigated topic in the last decades  \cite{narow}.  
The linear water wave scattering by variable depth (or with other name bottom topography) in the 
absence of a floating plate has been investigated 
by many authors. Two approaches have been developed. The first is analytical and the solution is derived in an almost
closed form \cite{port1995,stazik,port2000}. 
The second approach is numerical,  and developed by Liu and Liggett \cite{liu}, in which the boundary element method in a
finite region is coupled to a separation of variables solution in the semi-infinite outer domains. 
 For both the analytic and numerical approach the region of variable depth must be bounded.
Wave scattering by a floating elastic plate on water of variable bottom topography was treated 
by Wang and Meylan \cite{wang} in 2002.
Using Zakharov integral equation approach, a pair of coupled non-linear evolution
equations are derived for two co-propagating weakly non-linear gravity wave packets
over finite depth fluid \cite{chow}.
The newest results about disperse shallow water waves can be found in the monograph of Khakimzyanov {\it{et al.}}  \cite{disp}. 

We apply the following well-known self-similar Ansatz for the dynamical variables:  
\begin{eqnarray}
\zeta(x,y,t) = t^{-\alpha} f(\eta), \hspace*{3mm} 
u = t^{-\gamma} g(\eta),   \hspace*{3mm} 
v = t^{-\delta} h(\eta),   
\label{ansatz}
\end{eqnarray}
with the new reduced variable of $\eta = \frac{x+y}{t^{\beta}}$. 
All the exponents $\alpha,\beta,\gamma,\delta $ are real numbers. (Solutions with integer 
exponents are called self-similar solutions of the first kind, non-integer exponents 
generate self-similar solutions of the second kind.) 
This Ansatz is one kind of reduction mechanism where the original PDE system is reduced to an Ordinary Differential Equation (ODE) system. 
Unfortunately, both initial and the boundary problems become undefined. 
The obtained results can fulfil some kind of well-defined initial and boundary problems only via fixing their integration constants 
during the integration of the obtained ODE system.
The shape functions $f,g,h$ should be continuous functions and will be 
evaluated later on. (For first order Euler-type PDEs no continuous higher order derivatives of the shape functions are needed, therefore shock waves 
solutions, solutions with jumps or with singularity may occour.)  \\
The logic, the physical and geometrical interpretation of the Ansatz were exhaustively analyzed in all our 
former publications \cite{imre1,imre2,imre3,imre4, imre_book} therefore we neglect it.  
Except for some extreme cases positive exponents - now $\alpha,\beta,\gamma$ and $\delta$ mean physically 
reasonable and well-behaving dispersive solutions which decay and spread out in time. 
The exponent $\beta$ has a direct connections to the spreading velocity of the solution. 
The $\alpha, \gamma, \delta = 0$  case occupy a special place in 
our analysis and could mean physically relevant solutions without any temporal decay, and in this sense 
these are similar to solitions. (We cannot describe real solitons with our Ansatz, because that would mean zero dispersion an decay as well.)  \\
The big advantage of the  self-similar Ansatz of (\ref{ansatz}) that it directly gives us the Gaussian (or fundamental) solutions of 
the regular diffusion (or heat conduction) equation where the above mentioned physical meaning of the exponents are clear to identify. 
Applying this kind of Ansatz to any kind of non-linear PDE system gives us clear informations about the dispersive properties of the investigated 
phenomena. Solutions with negative exponents usually mean divergent solutions which will be mentioned later and 
might represent tsunamis or rough waves.   

Now we have to make a case studies for different well defined 
bed functions $\tilde{l}(x,y)$ or $l(x,y,t)$. 
Such kind of an analysis was performed with the Kardar-Parisi-Zhang (KPZ) interface growing equation where the effect of 
the different noise terms was investigated and numerous analytic solutions were derived \cite{kpz1}. 
We used the Maple 12 mathematical software to evaluate the analytic solutions of Eq. 
(\ref{eq0:aa1} - \ref{eq0:aa3}) for different seabed functions from now on. 
\subsubsection{The constant seabed function} 
Let's start with the simplest seabed function, namely with the $l(x,y,t) = d$ case where $d < 0$ is a real number. 
The obtained ODE system reads
 \begin{subequations}\label{eq1:system3}
\begin{empheq} {align}
      -\alpha f - \eta f'  + d g' + d h'  &    =  0, \label{eq1:aa1}
  \\  (-\alpha g -  \eta g')d  +  10 d f'   &    =  0,  \label{eq1:aa2}
  \\ (-\alpha h - \eta h')d  + 10 d f'   &    =  0,  \label{eq1:aa3}
\end{empheq}
\end{subequations}
where the prime means derivation in respect to the variable $\eta$ and 
the corresponding self-similar exponents are 
\eq
\alpha = \gamma = \delta = \textrm{ arbitrary real number,} 
\hspace*{1cm} \beta = 1. 
\eqe
Note, that the system is symmetric in the two velocity coordinates. It is important to say,  
that all three exponents which are responsible for the temporal decay of the dynamical variables 
($\alpha,\delta $ and $\gamma$) are arbitrary which gives a relatively large freedom for the solutions. 
Usually, positive exponents mean physically reasonable non-explosive solutions for large times. 
The common exponent $\beta$ which is responsible for the spreading is positive which is a promising sign  
for possible solutions. Furthermore, because $\beta=1$, 
a construction $\eta\pm C$, where $C$ is a constant, yields 
$\frac{x+y \pm C \cdot t}{t}$, which in itself is a kind of decaying wave in time.   
 
Unluckily, there are no closed form analytic solutions available for all three dynamical variables of 
(\ref{eq1:aa1} - \ref{eq1:aa3}) for general $\alpha$ and d parameters. 
However, for the wave height the general solution can be formulated in the next closed form of: 
\eq
f(\eta) = c_1(\eta + 2\sqrt{5d})^{-\alpha} +  c_2(\eta - 2\sqrt{5d})^{-\alpha}.
\label{f_const_1d}
\eqe
\begin{figure}  
\scalebox{0.80}{
\rotatebox{0}{\includegraphics{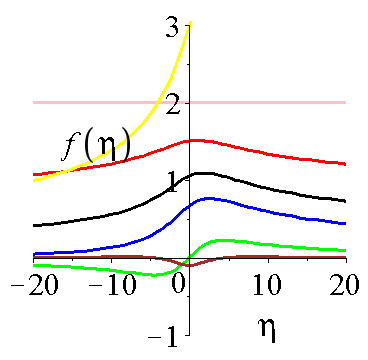}}}
\scalebox{0.60}{
\rotatebox{0}{\includegraphics{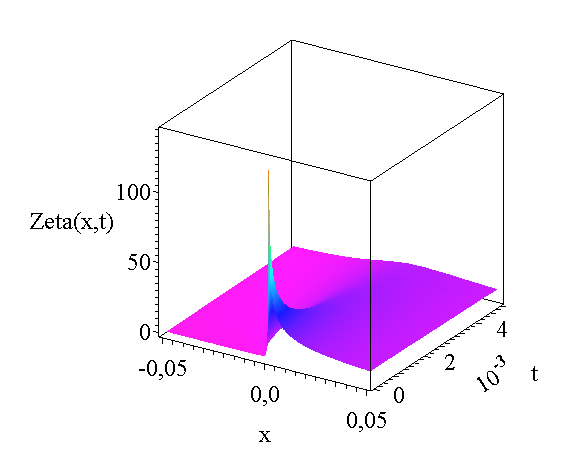}}}
\vspace*{-0.2cm} \\
  \hspace*{-1.4cm}  {\bf{a)}}   \hspace*{6cm}   {\bf{b)}} 
\caption{ a) The shape functions of the wave height (\ref{f_const_1d})  are presented  for various self-similar exponents. The yellow, pink,
 black, blue, red, green and brown curves are for $\alpha = -1/2, 0,
 1/3, 1/2, 2/3 , 1 $ and $ 2 $, respectively.  The integration constants $c_1,c_2$ were set to unity, and water depth $d = -1$. \\
b)The solution of the wave height function  of  $\zeta(x,y=0,t) = t^{-\alpha} f(x/t)$ for the parameter set o
f  $\alpha = 1/2, c_1 = 1, c_2 = 1$, the water depth is $ d = -1$.  } 
\label{kettes}        
\end{figure}   
Note, that for $\beta = 1$  we automatically have the $\eta = \frac{x}{C\cdot t}$ relation with  $C = 1$  (which is a real wave speed).
After a trivial algebraic step (\ref{f_const_1d}) can be reformulated to 
\begin{equation}
 f(\eta) =   c_1 t^{\alpha}[x + y +2\sqrt{5d} \cdot t]^{-\alpha}   + c_2 t^{\alpha} [x + y  - 2\sqrt{5d} \cdot t  ]^{-\alpha}  
  \sim   t^{-\alpha}  f(x+y  \mp  C t) 
\end{equation}
which are left and right running traveling waves with velocity of $ C = 2\sqrt{5d}$ and with  power-law time-varying amplitude of $t^{\alpha}$. 
Of course, the shape functions are now not the well-known sine or cosine functions but the power-law of the wave argument. 
This is an interesting and rare feature of the derived results. Such phenomena (when the original self-similar Ansatz leads to traveling-wave solutions with 
time-varying amplitude) has already occurred in our former studies. More than a decade ago we investigated a heat conduction model based on the Euler-Poisson-Darboux equation which is a "kind of time-dependent telegraph-type equation" with the usual self-similar Ansatz a
nd we found a solution which is a product 
of two (left-running and right-running) traveling waves with additional temporal decay \cite{imre_robi}.  
As further explanation we may say that, the original PDE system (\ref{eq0:system1}) is "so hyperbolic" (by that we mean 
a first-order system without any kind of  
additional dispersion) that even the self-similar Ansatz (which is very successful to obtain disperse and decaying 
solutions of parabolic systems - school example Gaussian solution of the diffusion equation) provides traveling wave 
solutions. Another interpretation could be the following time-decaying traveling 
wave solutions of hyperbolic systems can be found with the self-similar Ansatz due to it's internal structure.    
In other words this effect is a clear fingerprint of the in-depth entanglement of diffusion and wave phenomena.

 Figure (\ref{kettes})  a) presents the shape function of the wave height function for different $\alpha$ exponents. Note, that for $\alpha > 0$  the derived 
solutions have either a local minimum or maximum or both and an asymptotic decay to zero at large arguments.  For $\alpha < 0 $ the solution is divergent 
at large arguments.  Figure (\ref{kettes})  b)  shows the final wave height function $\zeta(x,y=0,t)$ for $\alpha = 1/2$ which has 
very sharp peak at the origin  at small times and distances and a steep decay. 
All other positive $\alpha$s show similar behaviour as well.   
The $\alpha = 0$ case means no wave at all, just a trivial constant function 
everywhere.  

For a better understanding we give the explicit forms of the function $f$ 
in case $d=-1$, $c_1=c_2=\kappa$ and some values of $\alpha$. 
If $\alpha = 1/2$, and $\kappa=1$, we have for the function $f$: 
\begin{equation}
f(\eta)= \frac{1}{(\eta+2\sqrt{5} i)^\frac{1}{2}} + 
\frac{1}{(\eta+2\sqrt{5} i)^\frac{1}{2}}
\end{equation} 
In this case if one assumes that 
\begin{equation}
(\eta+2\sqrt{5} i)^\frac{1}{2} = p + q i , 
\end{equation}
then 
\begin{equation}
(\eta-2\sqrt{5} i)^\frac{1}{2} = p - q i , 
\end{equation}
in the above evaluation. 
Considering 
\begin{equation}
\eta+2\sqrt{5} i = p^2 - q^2 + 2 pq i 
\end{equation}
then we have  
\begin{equation}
p^2= \frac{\eta+\sqrt{\eta^2+20}}{2}, q=\frac{\sqrt{5}}{p}
\end{equation}
Consequently the formula for $\alpha=1/2$: 
\begin{equation}
f(\eta)=\frac{1}{p+q \cdot i}+\frac{1}{p - q \cdot i}= 
\frac{2p}{p^2+q^2} .
\end{equation}
If we insert value for $p$ one gets 
\begin{equation}
f(\eta)= \frac{2 \sqrt{\frac{\eta+\sqrt{\eta^2+20}}{2}} }{ \frac{\eta+\sqrt{\eta^2+20}}{2} + \frac{5}{\frac{\eta+\sqrt{\eta^2+20}}{2}} } .
\end{equation}

If $\alpha=1$ is inserted, one gets 
\begin{equation}
f(\eta) = \kappa \frac{2\eta}{\eta^2+20} , 
\end{equation}  
in case $\alpha=2$ 
\begin{equation}
f(\eta) = \kappa \frac{2(\eta^2-20)}{(\eta^2-20)^2 + 80} . 
\end{equation}

\begin{figure} 
\scalebox{0.5}{
\rotatebox{0}{\includegraphics{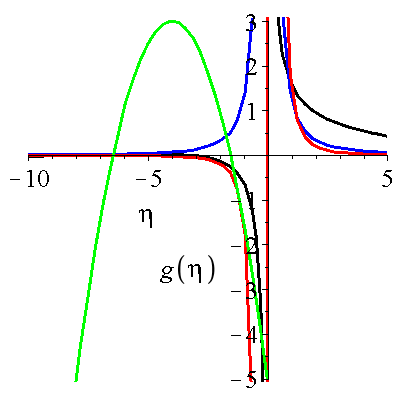}}}  \hspace*{0.5cm}
\scalebox{0.5}{
\rotatebox{0}{\includegraphics{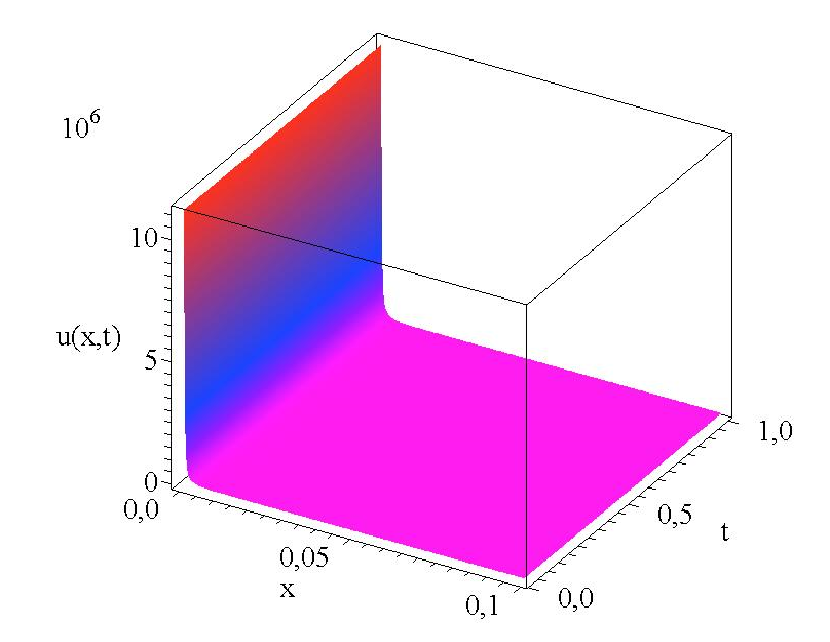}}} \\
  \hspace*{-1.8cm}{\bf{a)}}   \hspace*{6.5cm}   {\bf{b)}} 
\caption{ a) The sharp function of the x velocity component $g(\eta)$ for four  different $\alpha$ exponents for the constant seabed function. 
The black, blue, red and green line correspond to $	 1,2,3$ and $-2$ values, respectively.  The integral constants $c_1 = c_2 = c_3 = 1$ and $d = -1$.  \\ 
b) The velocity projection $u(x,y=0,t)  = t^{-\gamma}g(\eta) $ for $\gamma = 2$ other integration constants and the water depth remain the same.  } 
\label{harmas}        
\end{figure}

Note, that our PDE or even the obtained ODE system is symmetric in the two velocity coordinates, so it is enough to evaluate and analyze one of them.  
For the velocity variables the general solutions can be formulated, but contain an integral which can be evaluated in closed forms for 
given integer $\alpha$ values only.   
For $\alpha > 0 $ the solutions are proper rational fractions, having singularity in the origin and zero asymptotic values in infinity, 
for  $\alpha \le 0 $ the results are polynomials. We give two examples:  
 \begin{subequations}\label{eq:system}
\begin{empheq} {align}
   \alpha = 1, \hspace*{1cm} g =  \frac{10 c_2 \eta + 10 c_3 + c_1 \eta^2 + 20 c_1}{(\eta^2+20)\eta},  \\
    \alpha = -2, \hspace*{1cm}  g =  \left(  \frac{c_2}{2}  - c_1 \right)\eta^2   - 20c_3\eta - 5c_2,    
    \label{lin}
\end{empheq}
\end{subequations}
On Figure (\ref{harmas})  a) we present the velocity shape function $g(\eta)$ for four different $\alpha$ exponents.  
The projection of the velocity function  $u(x,y=0,t)  = t^{-\gamma}g(\eta) $ for $\gamma = 2$ is visualized on Figure (\ref{harmas}) b).  
Note the power law decay for large distances.  The wave velocity as a relevant dynamical variable of the system shows no extra features.   
 
\subsubsection{The linear seabed function} 
The next choice is the linear case. We have distinguish two different cases and have to make a separation
 \begin{subequations}\label{eq33:system3}
\begin{empheq} {align}
 \tilde{l}(x,y) &= - (ax + by + c),  \label{eq33:aa1} \\ 
 l(x,y,t) &= - (ax + by + c)/t^{\beta} = - \eta,    \label{eq33:aa2} 
\end{empheq}
\end{subequations} where the negative sign were taken from geometrical reasons, we want to have the seabed in the negative region, 
$a,b$ and $c$ are just free positive real numbers which are responsible for the slope of the seabed and the depth in the origin.  
The first case means a rigid sea-bed which does not change it's shape during the wave motion. Probably this model is more relevant for deeper
water where the sea bottom is rigid.  
In our forthcoming analysis we have to introduce the following identity  
\eq
 \tilde{l}(x,y) = - (ax + by + c) = -\frac{ax + by + c }{t^{\beta}} \cdot t^{\beta} = - \eta \cdot t^{\beta}. \\ 
 \eqe  

The second case  (\ref{eq33:aa2}) represents a physical situation where the sea-bed function is continuously modified during the 
corresponding wave propagation which is feasible in very shallow water and loose ground soil like sand.
Note, that the $t^{\beta}$ factor should play a relevant role in the time asymptotic. 

Let's analyze the time-independent case first. After some simple algebraic manipulations we can derive the ODE system of 
 \begin{subequations}\label{eq2:system3}
\begin{empheq} {align}
      -\alpha f' - 2\eta f'  -  g'\eta - ga - h'\eta - hb  &    =  0, \label{eq2:aa1}
  \\  ([-\alpha-1]g -  2\eta g')\eta  -  10 \eta f'   &    =  0,  \label{eq2:aa2}
  \\ ([-\alpha-1]h - 2\eta h')\eta  - 10 \eta f'   &    =  0,     \label{eq2:aa3}
\end{empheq}
\end{subequations}
with exponents of 
\eq 
 \alpha +1  = \delta = \gamma =\textrm{ arbitrary real number}, \hspace*{1cm} \beta = 2. 
\eqe 
Note, the slight difference compared to the constant seabed function.  
The $\beta = 2$ exponent which is responsible for the "spreading" is favourable high, in regular diffusion processes it is usually just $1/2$.   
There are no general formulas available for all three dynamical variables for arbitrary parameters $a,b$ and for the $\alpha$ exponent.  
The exception is the shape function of the wave height which is the following: 
 \begin{eqnarray}
 &f =   {c_{1}} \cdot  \>  _{2}F_{1} \left(\frac{\alpha}{2}, \frac{\alpha+1}{2}; \frac{a+b}{2};  \frac{\eta}{5} \right) +  \nonumber \\ 
&  c_{2} \cdot    
 \eta^{1- \left[\frac{a+b}{2} \right]}  \cdot _{2}F_{1} \left(\frac{\alpha-b-a}{2}+1, 
\frac{3-b-a+ \alpha}{2}; 2-\frac{a+b}{2};  \frac{\eta}{5} \right)
\label{time_indep}
 \end{eqnarray}
where $ _{2}F_{1}\left[\cdot,\cdot;\cdot;\cdot\right]$ are the hypergeometric functions \cite{NIST}. 
The hypergeometrical function is defined for $ |z| < 1$ by the power series of 
\begin{eqnarray}
_{2}F_{1} \left(a, b; e; z \right) = \sum_{n=1}^{\infty} \frac{ (n)_n (b)_n}{  (e)_n} \frac{z^n}{n!} = 
1 + \frac{ab}{e} \frac{z}{1!} + \frac{a(a+1)b(b+1)}{e(e+1)} \frac{z^2}{2!} + ... 
\end{eqnarray}
where $(q)_n$ in the (rising) Pochhammer symbol, which is defined by: \\ 
\begin{eqnarray}
    (q)_n= 
\begin{cases}
      1, & \text{if } n = 0,\\
     q(q+1) \cdot \cdot \cdot (q+n-1),              & \text{if } n > 0. 
\end{cases}
\end{eqnarray}

It is clear to see from the infinite power series definition of the hypergeometric function using the 
Pochhammer symbols, that the function is undefined (or infinite) if the third parameter equals a non-positive integer. 
On the other side the infinite series terminates if the first two parameters of the hypergeometric functions are non-positive integers. 

It is a hopeless undertaking to give a general parameter study of the solution for arbitrary $\alpha, a$ and $b$. 
It is logically clear that larger parameters $a,b$ cause a steeper slope which affects the numerical values of the exponent. 
So lets, fix $a = b = 1$ and investigate the role of the self-similar exponent $\alpha$ 
now the former formula (\ref{time_indep}) is changed to 

 \begin{eqnarray}
 &f =    (\eta - 5)^{-\frac{1+ 2\alpha}{4}} \cdot   \left( c_1 \cdot  _{2}F_{1} \left[\frac{\alpha}{2}, \frac{\alpha+1}{2}; 
 \frac{2\alpha +1}{2};  \frac{5-\eta}{5} \right]  \cdot  \left[ \frac{5- \eta}{5} \right]^{\frac{1+ 2\alpha}{4}} 
 \right. \nonumber \\ 
 & \left.   +  c_{2} \cdot     _{2}F_{1} \left[\frac{2-\alpha}{2}, \frac{1- \alpha}{2};   \frac{3 - 2\alpha}{2} ; 
 \frac{5-\eta}{5} \right]  \cdot  \left[ \frac{5-\eta}{5} \right]^{\frac{3 - 2\alpha}{4}}     \right) 
\label{time_indep_ab1}.
 \end{eqnarray}

\begin{figure}  
\scalebox{0.55}{
\rotatebox{0}{\includegraphics{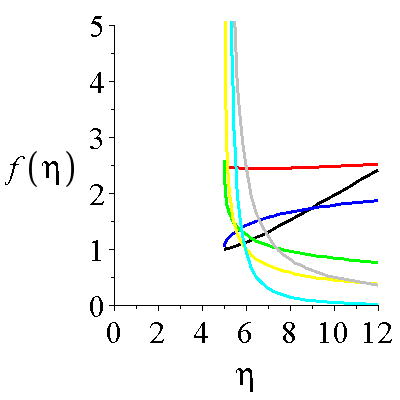}}}
\scalebox{0.55}{   
\rotatebox{0}{\includegraphics{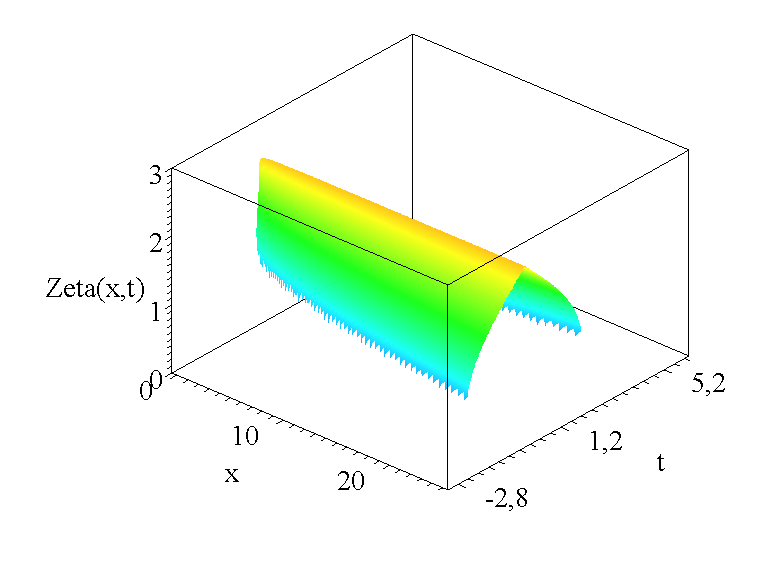}}}  \\
  \hspace*{1cm}{\bf{a)}}   \hspace*{5.0cm}   {\bf{b)}}  
\caption{ a)  The shape function of wave height $f(\eta)$ for various $\alpha$s for time-independent linear seabed function. 
The black, red, blue, green, yellow, grey  and cyan colours are for $\alpha = -1, -1/2, 0, 1/2, 1, 3/2$ and for 2. \\ 
b)  The the spatial and temporal dependence of the water wave height  
 $\zeta(x,y = 0,t)  $  for $\alpha = 0$ and for $ a = b = 1$ and $c = 0$.  
  } 
\label{negyes}        
\end{figure}

Figure (\ref{negyes})  a) presents  (\ref{time_indep_ab1})  for various $\alpha$s.  
It is clearly visible, that for $\alpha < 0$ all shape functions diverge at large argument $\eta$. 
For positive $\alpha$s the shape functions goes to zero at infinity. Note, the lower limit of the domain of all shape functions which lie uniformly at 
$\eta = 5 $ which is half of the gravitational acceleration parameter of the system. We got the usual features that 
negative exponents mean non-decaying solutions at large arguments. 

On the right side of Fig. (\ref{negyes}) the $y = 0$ projection of the wave height $\zeta(x,y=0,t)$ is presented for
$\alpha = 0$  which means that the global maximum is not changing in time.  The most important 
property of the solution is it's shape, which describes a continuously  widening ridge with a compact 
support in space. The function together with the first spatial and temporal derivatives remain finite at the 
border of the domain.  The value and the spatial position of the maximum is not changing in time,  
therefore it cannot be interpreted as any kind of traveling wave. 

\begin{figure} 
\scalebox{0.58}{
\rotatebox{0}{\includegraphics{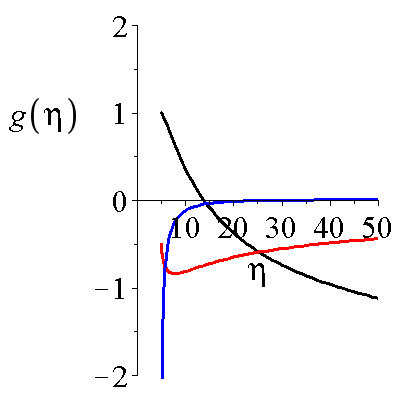}}}  \hspace*{0.5cm}
\scalebox{0.58}{
\rotatebox{0}{\includegraphics{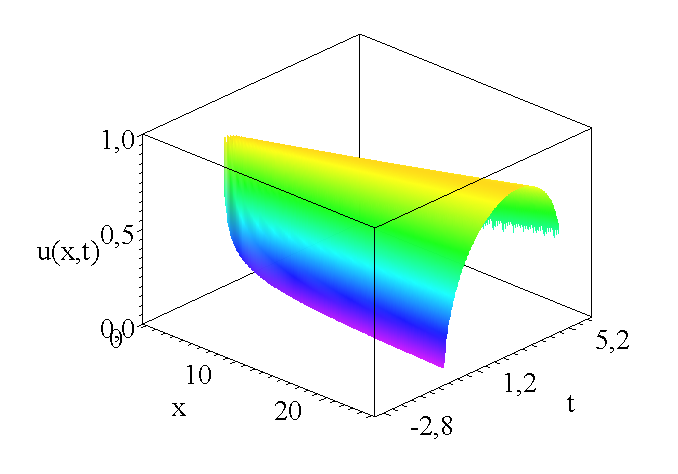}}} \\
  \hspace*{-1.8cm}{\bf{a)}}   \hspace*{6.5cm}   {\bf{b)}} 
\caption{ a) The shape function of the x velocity component $g(\eta)$ for four  the above given 
three  $\alpha$ exponents for the time-independent 
 seabed function. 
The black, red and blue lines correspond to $	\alpha = -1,0$ and $1$ values, respectively.  
The integral constants $c_1 = c_2 = c_3 = 1$  the geometrical  constants are $a = b = 1$ and $c = 0$.  \\ 
b) The spatial and temporal dependence of velocity component $u(x,y=0,t)$ for $\alpha = 0$   all other constants remain the same.  } 
\label{otos}        
\end{figure}

\begin{figure}  
\scalebox{0.55}{
\rotatebox{0}{\includegraphics{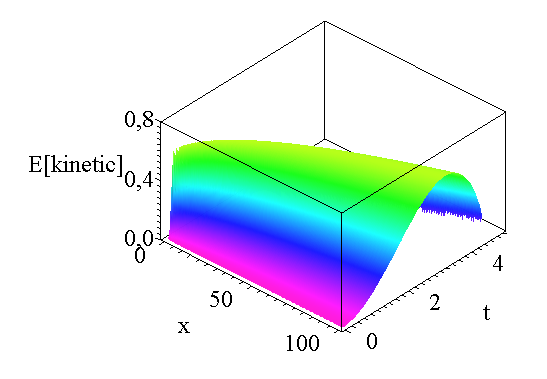}}}
\caption{   The "Kinetic enegy-like property" of the wave according to Eq. 
(\ref{Ekin}) with all numerical parameters given above.} 
\label{haromB}        
\end{figure}

There is no general formula available for arbitrary $\alpha$ for the velocity shape function, however for fixed and 
well-defined $\alpha$s closed formulas can be evaluated.  We present some shape functions for $a = b = 1$ 
  \begin{subequations}\label{eq:system22}
\begin{empheq} {align} 
    \alpha & = -1, \hspace*{1cm}  g =  c_2 + \frac{5c_3}{\sqrt{\eta-5}} - c_3\sqrt{5}arctan\left( \frac{\sqrt{5\eta-25}}{5}\right) - 
\frac{25c_3}{\sqrt{\eta - 5\eta}} -c_1,   \\
    \alpha & = 0, \hspace*{1cm}  g =   -\frac{5c_3 \eta^{3/2} -25c_3\sqrt{\eta} +c_1\eta \sqrt{5\eta - 25} }{\sqrt{5\eta - 25\eta}\eta^{3/2} }   \\ 
   \alpha & = 1, \hspace*{1cm}  g =  \frac{-5c_3}{\eta \sqrt{\eta -5}} + \frac{c_2}{5}   
    \label{lin22}
\end{empheq}
\end{subequations}
For other e.g. non-integer $\alpha$s we get expressions which contain an additional integration with some extra constants.
These integrals can be evaluated only when all constants are fixed. 
 On Fig. (\ref{otos}) a) we can see  the three analytic velocity shape functions of  (\ref{eq:system22}).  Note, that all 
three functions are defined for $\eta > 5$, only. 
To have a complete picture Fig. (\ref{otos}) b)  shows the velocity distribution $u(x,y,=0,t)$ for the $\alpha = 0$ exponent which means that the 
global maximum has temporal decay again. The general feature of the wave velocity function is 
very similar to the former wave height function. It is a continuously widening ridge with a compact support is space again.  The function together with the corresponding first spatial and temporal derivatives 
has finite values the the border of the domain. However the spatial position moves in time which means that it is a traveling wave solution.  
To have a much picturesque presentation of how the water wave behaves we present a plot which shows the 
"kinetic-energy like" property of the wave namely the 
\eq
E_{kin} \propto  \frac{\zeta(x,y=0,t) \times u(x,y=0,t)^2}{2}. 
\label{Ekin}
\eqe
Our argumentation is the following we do not have the density as a 
dynamical variable of the model, but we think that the true height of the 
waver wave is proportional to the mass of the wave, therefore the presented 
quantity gives us a physically acceptable hint of the wave. 
As we can see the domain of the height and the velocity functions are slightly different, a physical true wave is 
only present when the mass (or now the wave height) and the velocity are different to zero.  
Figure \ref{haromB} presents the distribution of Eq. (\ref{Ekin}). 
Note that the shape of the wave describes a true finite traveling wave in space 
and time with constant amplitude.  Due to the $\beta = 2$ self-similar exponent (which is responsible for the 
dispersion ratio of the dynamical variables), the 
obtained wave has strong spreading, therefore despite of the constant 
maximum  our wave is not a solitary wave.  The rising edge of the wave has 
a finite non-zero time derivative at the boarder of the domain and a falling edge has a zero derivative at $ t = 0$. 
Last we have to note, that the role of depth of the seabed $c$ does not explicitly appear in the shape functions, 
however it does effect the final solutions and  can be visualized on the final 
space and temporal dependences of the wave height and velocity  
(Fig. 4b and Fig. 5b)  or in Fig. 6. The larger the water depth is the broader 
the corresponding physical quantity.  Larger water depth does not modify the quality of the solutions just make them more disperse. 
 So with a fixed numerical value (of eg. $c = 40$) and a maximal 
wave height of 3 we have a good approximation of deep water phenomena.  
In general it is a remarkable fact that our self-similar Ansatz is capable to describe 
traveling wave pattern in a first-order hyperbolic PDE system.   
 \\ 


For our last case for time-dependent linear seabed function  (\ref{eq33:aa2}) we have a bit different reduced ODE system of 
 \begin{subequations}\label{eq3:system3}
\begin{empheq} {align}
  -\alpha f' - \eta f'  -  g'\eta - ga -  h'\eta - hb  &    =  0, \label{eq3:aa1}
  \\  -(-\alpha g -  \eta g')\eta + g\eta  -  10 \eta f'   &    =  0,  \label{eq3:aa2}
  \\ -(-\alpha h - \eta h')\eta +h\eta - 10 \eta f'   &    =  0,     \label{eq3:aa3}
\end{empheq}
\end{subequations}
with the exponents of 
\eq 
 \alpha  = \delta = \gamma =\textrm{ arbitrary real number}, \hspace*{1cm} \beta = 1. 
\eqe 
Note, that all the exponents which are responsible for dissipation can have 
arbitrary values as before (which allow exploding or decaying solutions or solutions with constant 
maximum in time). The only difference is the numerical 
value of $\beta$ which is now unity.   

\begin{figure}  
\scalebox{0.6}{
\rotatebox{0}{\includegraphics{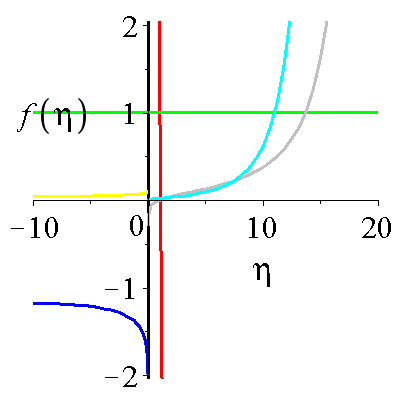}}}
\scalebox{1.}{   
\rotatebox{0}{\includegraphics{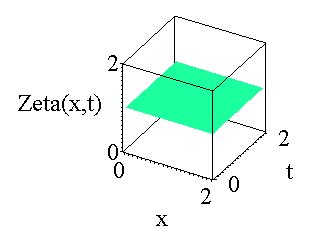}}}  \\
 \hspace*{1cm}{\bf{a)}}   \hspace*{5.0cm}   {\bf{b)}}  
\caption{ a)  The shape function of wave height $f(\eta)$ for various $\alpha$s for time-dependent linear seabed function. 
The black, red, blue, green, yellow, gray  and cyan colours are for $\alpha = -2, -1,-1/2, 0, 1/2, 1$ and for 2, the 
geometrical constants are $a = b =1$ and $c = 0$   . \\ 
b)  The the spatial and temporal dependence of the water wave height  
 $\zeta(x,y = 0,t)  $  for $\alpha = 0$.  } 
\label{hatos}        
\end{figure}

As in the previous case, no closed-form analytic solutions exist for all dynamical variables for arbitrary parameters, except for the  
 wave height 
 \begin{eqnarray}
f =  &(\eta + 20)^{\left(\frac{a+b}{4} -\alpha -1\right)}  \cdot \left( -\frac{\eta}{20} - 1 \right)^{\frac{a+b}{4}-\alpha} \cdot \left(  
c_1 \cdot   _{2}F_{1}\left[-\alpha , 1-\alpha ; 2 - \frac{a+b}{2}; - \frac{\eta}{20} \right ]\eta^{\left(1- \frac{a+b}{2}\right)}  +  \right. \nonumber \\ 
& \left. + c_2  \cdot    {2}F_{1}\left[  \frac{a+b}{2}- \alpha,  \frac{a+b}{2} -\alpha -  1     ; \frac{a+b}{2}   ; - \frac{\eta}{20}  \right]  \right);
\label{time_dep}
 \end{eqnarray}
where the two $ _{2}F_{1}\left[\cdot,\cdot;\cdot;\cdot\right]$s are still the hypergeometric functions \cite{NIST}. 
The complete general parameter study looks hopeless. Let's fix $a = b =1$ first, now the solution looks a bit more transparent 

\begin{eqnarray}
f =  &(\eta + 20)^{-\left(\alpha + \frac{1}{2}\right)}  \cdot \left(  
c_1 \cdot   _{2}F_{1}\left[\alpha , \alpha+1 ; 1 + 2\alpha ; 1- \frac{\eta}{20} \right ]\cdot [1-\frac{\eta}{20}] \eta^{\left(\alpha +\frac{1}{2}\right)}  +  \right. \nonumber \\ 
& \left. + c_2  \cdot    {2}F_{1}\left[ - \alpha,  1 -\alpha; 1-2\alpha;1 - \frac{\eta}{20}  \right] \cdot [1-\frac{\eta}{20}]^{\frac{1}{2}-\alpha}    \right).
\label{time_dep_1}
\end{eqnarray}
For some relevant rational $\alpha$ values the shape function becomes 
more simple like: 
    \begin{subequations}\label{eq:system33}
\begin{empheq} {align} 
    \alpha & = -1, \hspace*{0.8cm}  f =    c_2 +c_3\eta - 20c_3 ln(\eta)   \\
    \alpha & = 0, \hspace*{1cm}  f =   c_2   \\ 
   \alpha & = \frac{1}{2}, \hspace*{1cm}  f =   \frac{c_2 LegendreP\left( -\frac{1}{2},1, 1-\frac{40}{\eta}  \right)+ c_3LegendreQ\left( -\frac{1}{2},1, 1-\frac{40}{\eta}  \right)  }{\sqrt{\eta} \cdot \sqrt{\eta - 20}}  \\ 
\alpha& = 1 \hspace*{1.1cm}  f = \frac{-c_2 \eta + 20 c_2 ln(\eta) +c_3}{(\eta - 20)^2}. 
    \label{lin33}
\end{empheq}
\end{subequations} 
Where LegendreP($\cdot,\cdot;\cdot$) and  LegendreQ($\cdot,\cdot;\cdot$) are the regular and irregular Legendre functions \cite{NIST}. 
The left part of Figure (\ref{hatos}) presents the shape function of the water height (\ref{time_dep_1}) for some $\alpha$ exponent values.   
If $\alpha $ is equal with $-1$ and $-2$ the graphs (the black and red curves) are almost vertical lines. 
Zero alpha means a constant shape functions which is quite reasonable. 
For $|\alpha| = 1/2$ the shape functions have finite values in the origin, can be expressed with Legendre functions 
and are defined for negative arguments only.  For positive integer $\alpha$s the functions have at asymptote at $\eta = 20$. 
On the right side of Fig.  (\ref{hatos})  the water height function is plotted for the $\alpha = 0$ exponent. 
The solution is the trivial constant wave height function for every time and space point.  

\begin{figure} 
\scalebox{0.64}{
\rotatebox{0}{\includegraphics{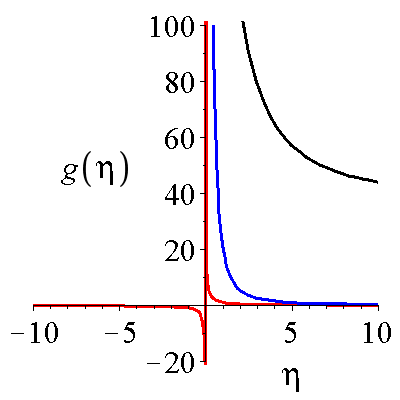}}}  \hspace*{0.5cm}
\scalebox{0.64}{
\rotatebox{0}{\includegraphics{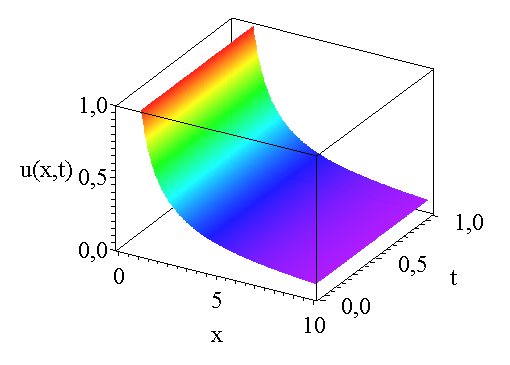}}} \\
  \hspace*{-1.8cm}{\bf{a)}}   \hspace*{6.5cm}   {\bf{b)}} 
\caption{ a) The shape function of the x velocity component $g(\eta)$ for three $\alpha$ exponents for the time-dependent 
 seabed function. 
The black, red and blue lines correspond to $	\alpha = -1,0$ and $1$ values, respectively.  The integral 
constants are $c_1 = c_2 = c_3 = 1$ and the geometrical are the same as above.   \\ 
b) The spatial and temporal dependence of velocity component $u(x,y=0,t)$ for $\alpha = 0$.  } 
\label{hetes}        
\end{figure}

Now comes the the analysis of the velocity field. 
As before, there are no analytic solutions available for the entire parameter space. For fixed a and b steepness 
there are closed formulas available for integer $\alpha$s only. Unfortunately, for rational $\alpha$s the solutions contain and additional 
formal integration which cannot be evaluated with analytic means. 
The shape functions for the x component of water velocity are the following for the most relevant  three integer $\alpha$s:
 \begin{subequations}\label{eq:system44}
\begin{empheq} {align} 
    \alpha & = -1, \hspace*{0.cm}  g =   \frac{200 c_2}{\eta}  + 10c_3 ln(\eta) + c_1,   \\
    \alpha & = 0, \hspace*{0.3cm}  g =   \frac{c_2}{\eta},   \\ 
   \alpha & = \frac{1}{2}, \hspace*{0.31cm}  g =   \frac{-600c_2\eta + 10c_2\eta ln(\eta) + 20c_3\eta   + c_1\eta^2 - 
40c_1\eta + 200(2c_1 + 40c_2 - c_3)  }{(\eta - 20)^2\eta^2}. 
    \label{lin44}
\end{empheq}
\end{subequations} 
For the shake of completeness Figure (\ref{hetes}) a) presents these three shape functions and Figure (\ref{hetes}) b) shows the 
 projected velocity field for $\alpha = 0$.    
We tried to evaluate and plot Eq. (\ref{Ekin}) for various $\alpha$s, unfortunately we cannot find and reasonable function which 
could be interpreted as any kind of finite water wave.  
So there are analytic solutions available, but these are out of any physical interest. \\ 
\subsection{Traveling wave analysis} 
Lastly, we try to analyze the original PDE system of Eq.  (\ref{eq0:system1}) with the traveling wave Ansatz.  
Numerous kind of traveling wave solutions exist for numerous water wave 
equations. One of them is the Wilton ripple which is a type of periodic traveling wave solution of the full water wave problem incorporating the effects of surface tension \cite{olga}.

For our system we take the form of the traveling wave as  
\begin{eqnarray}
\zeta(x,t) = f(x + c\cdot t)  =  f(\eta), \hspace*{3mm} 
u =  g(\eta),   \hspace*{3mm} 
v = h(\eta)  
\label{ansatz2}
\end{eqnarray} 
where c is the velocity of the wave.   
The obtained ODE system is slightly different from (\ref{eq1:aa1} - \ref{eq1:aa3})   

\begin{subequations}
\begin{empheq} {align}
      c f'   + g'l + gl' + h'l + hl'   &    =  0, \label{eq:aaa1}
  \\  c(g'l  +  gl')  + Glf'   &    =  0,  \label{eq:aaa2}
  \\  c(h'l  +  hl') + Glf'   &    =  0.    \label{eq:aaa3}
\end{empheq}
\label{eq2:system3}
\end{subequations}
Note, that for generality we consider that the sea bed 'l'  has the dependence of the traveling wave argument $\eta$. 
The physical meaning of such a space and time-dependent surface is of course questionable, and can be a scope of a different study.      
Here we just present the mathematical obtained results. According to Maple12 the undetermined system of (\ref{eq2:system3})  
has three different kind of solutions: 
\begin{eqnarray} 
&1)& \hspace*{0.5cm}  f(\eta) = c_1,  \hspace*{1.0cm}  l(\eta) =0, \hspace*{1cm}    \textrm{ and  g and h are arbitrary functions},  \\ 
&2)&  \hspace*{0.5cm} f(\eta)  \hspace*{0.3cm} \textrm{is arbitrary,}  \hspace*{0.3cm}    l(\eta) = \frac{c^2}{20}, \hspace*{0.3cm}  
 g(\eta) =-\frac{10 f(\eta)}{c} + c_1,  h(\eta) =-\frac{10 f(\eta)}{c} + c_2,   \\ 
&3)&  \hspace*{0.5cm}  f(\eta) = c_3,   \hspace*{1cm}    \textrm{ l is arbitrary function}, \hspace*{0.3cm}  h(\eta) =\frac{c_2}{l(\eta)}, \hspace*{0.3cm}  g(\eta) =\frac{c_1}{l(\eta)}, 
\end{eqnarray} 
where $c_1,c_2,c_3$ are the usual integration constants. 
Note, that for the third kind of solution the wave velocities which are inverse proportional to the sea-bed function.  
Above deep lying sea-beds  the velocity is small which can be understood as the direct manifestation of the Bernoulli principle.  
\section{Summary and Outlook }
We investigated a hydrodynamic system which is capable to describe water wave propagation over a variable water depth.  
The question of constant and linear sea-bed functions were addressed and analyzed. We found general analytic formulas for the 
spatial and time dependent wave heights, unfortunately for the velocity functions no general formulas are available. 
If all the parameters are fixed then closed formulas can be derived as well which contain various hypergeometric functions with 
non-trivial arguments. As second case the water waves are investigated above a time-independent (we may say) linear seabed function. 
We found that dispersive traveling wave solutions may exist with constant wave height in time. 
As a third case, we considered a time dependent linear seabed function and tried to 
find physical water waves with reasonable velocities and wave height. 
Unfortunately no such waves can be found among the mathematically existing solutions.   
 Finally, we investigated the traveling soultions and found not so interesting trivial solutions of the problem.  \\ 
In a former independent study we investigated the self-similar solutions of rotating and stratified fluids \cite{rot} and 
presented analytic results. We think that thanks to Nathan Paldor's monograph \cite{rot_shallow} investigating a 
rotating shallow water fluid systems could be a natural generalization of our present study. 
Consideration of viscosity could be a way of generalization of course. 
\section{Authors Contributions}
The corresponding author (I.F. Barna) provided the original idea of the study, performed the analytic calculations, 
created the figures and wrote 
some part of the manuscript. The second author (M. A. Pocsai) checked the spelling, improved the language of the final manuscript.  
The third author (L. M\'aty\'as) evaluated certain formulas, checked all the 
analytic calculations, the spelling, improved the language of the 
final manuscript and collected large part of the cited references as well. The authors discussed the manuscript on a regular weekly basis. 
 
\section{Acknowledgment}
One of us (I.F. Barna) was supported by the NKFIH, the Hungarian National Research 
Development and Innovation Office. 
\section{Conflicts of Interest}
 The authors declare no conflict of interest.
\section{Data Availability}
The data that supports the findings of this study are all available within the article. 
At the request of readers, the appropriate Maple files are available. 

\end{document}